\documentclass[aps,showpacs,floatfix,twocolumn,amsmath,amssymb,prd]{revtex4}
\usepackage{graphicx}
\usepackage{subfigure}
\usepackage{multirow}
\usepackage[citecolor=blue,colorlinks=true,urlcolor=blue,linkcolor=blue]{hyperref}
\begin{document}

\def\fm {\,{\rm fm}}
\def\MeV {\,{\rm MeV}}
\def\GeV {\,{\rm GeV}}

\title{The $\Delta_{mix}$ parameter in the overlap on domain-wall mixed action}
\author{M.~Lujan$^{1}$, A. Alexandru$^{1}$, 
T.~Draper$^{2}$, W. Freeman$^{1}$, M.~Gong$^{2}$, 
F.X. Lee$^{1}$, A.~Li$^{3, 4}$, K.F.~Liu$^{2}$, N. Mathur$^{5}$   \\
{\bf ($\chi$QCD Collaboration)}
\vspace*{0.5cm} }
\affiliation{$^1$Physics Department, The George Washington University, Washington, DC 20052, USA \hfill\\
$^2$Department of Physics and Astronomy, University of Kentucky, Lexington, KY 40506, USA \\
$^3$Department of Physics, Duke University, Durham, NC 27708, USA \\
$^4$Institute for Nuclear Theory, University of Washington, Seattle, 98195, USA\\
$^5$Department of Theoretical Physics, Tata Institute of Fundamental Research, Homi Bhabha Road, Mumbai 400005, India}

\begin{abstract}
\begin{center} {\bf{Abstract}} \end{center}
A direct calculation of the mixed-action parameter 
$\Delta_{mix}$ with valence overlap fermions on a domain-wall fermion sea is
presented. The calculation is performed on four ensembles of the 2+1-flavor
domain-wall gauge configurations: $24^3 \times 64$ ($a m_l= 0.005$, $a=0.114\fm$) and
$32^3 \times 64$ ($a m_l = 0.004, 0.006, 0.008$, $a=0.085\fm$). For pion masses
close to $300\MeV$ we find \hbox{$\Delta_{mix}=0.030(6)\GeV^4$} at $a=0.114\fm$ and $\Delta_{mix}=0.033(12)\GeV^4$ at $a=0.085\fm$.
The results are quite independent of the lattice spacing and they are significantly smaller than the results 
for valence domain-wall fermions on Asqtad sea or those of valence overlap fermions on clover sea. 
Combining the results extracted from these two ensembles, we get $\Delta_{mix}=0.030(6)(5)\GeV^4$,
where the first error is statistical and the second is the systematic error
associated with the fitting method.
\end{abstract}

\maketitle

\section{Introduction}

Mixed action approaches have been studied by several groups,
such as Domain-wall fermion (DWF) valence on Asqtad fermion sea~\cite{Edwards:2005ym}, overlap valence on DWF sea~\cite{Allton:2006nu}, overlap valence on 
clover sea~\cite{Durr:2007ez}, and overlap valence
on twisted-mass fermion sea~\cite{Cichy:2009dy}. In view of the fact that it is numerically intensive to simulate chiral fermions
(DWF or overlap), it is deemed practical to use the cheaper fermion formulation for generating gauge configurations and
the more expensive fermion discretization for the valence as an expedient approach toward dynamical QCD simulations with chiral fermions.
Many current algebra relations depend only on the chiral properties of the
valence sector. The mixed action theory with different fermions for the valence and the sea is a generalization of the
partially quenched theory with different sea and valence quark masses.
The mixed action partially quenched chiral perturbation theory (MAPQ$\chi$PT) has been developed for
Ginsparg-Wilson fermions on Wilson sea~\cite{Bar:2003mh} and staggered sea~\cite{Bar:2005tu}, and has been worked out for
many hadronic quantities to next-to-leading order (NLO), such as pseudoscalar masses and decay 
constants~\cite{Bar:2003mh,Bar:2005tu,Chen:2007ug},
isovector scalar $a_0$ correlator~\cite{Golterman:2005xa,Prelovsek:2005rf,Aubin:2008wk,WalkerLoud:2008bp},
heavy-light decay constants~\cite{Aubin:2005aq}, and baryon masses~\cite{Tiburzi:2005is, WalkerLoud:2008bp}.

In the mixed action chiral perturbation theory with chiral valence fermions, it is shown~\cite{Bar:2003mh} that to
NLO there is no $\mathcal{O}(a^2)$ correction to the valence-valence meson mass due to
the chiral symmetry of the valence fermion. Furthermore, both the chiral Lagrangian and the chiral extrapolation
formulas for hadron properties to the one-loop level (except $\theta$-dependent quantities) are independent of the
sea fermion formulation~\cite{Chen:2006wf}. The LO mixed-action chiral Lagrangian involves only one more term with
$\mathcal{O}(a^2)$ discretization dependence which is characterized by a low energy constant $\Delta_{mix}$.
The LO pseudoscalar meson masses for overlap valence and DWF sea are given as
\begin{eqnarray}   \label{eq:dd}
m_{vv'}^2 &=& B_{ov}(m_v +m_{v'}), \nonumber \\
m_{vs}^2 &=& B_{ov}m_v + B_{dw}(m_s + m_{res}) + a^2\Delta_{mix}, \nonumber \\
m_{ss'}^2 &=& B_{dw}(m_s + m_{s'} + 2 m_{res}),
\end{eqnarray}
where $m_{vv'}/m_{ss'}$ is the mass of the pseudoscalar meson made up of valence/sea quark and antiquark.
$m_{vs}$ is the mass of the mixed valence and sea pseudoscalar meson. Up to numerical accuracy, there is no
residual mass for the valence overlap fermion. The DWF sea has a residual mass $m_{res}$ which vanishes as
$L_S \rightarrow \infty$. $\Delta_{mix}$ enters in the mixed meson mass $m_{vs}$ as an
$\mathcal{O}(a^2)$ error which vanishes in the continuum limit. We should note that, unlike the partially
quenched case, even when the quark masses in the valence and sea match, the unitarity is still violated due
to the use of mixed actions. The degree of unitarity violation at finite lattice spacing depends on the size of $\Delta_{mix}$.

$\Delta_{mix}$ has been calculated for DWF
valence and Asqtad fermion sea which gives 
$\Delta_{mix} = 0.249(6)\GeV^4$ \cite{Orginos:2007tw} at $a=0.125\fm$, $0.211(16)\GeV^4$ at $a=0.12\fm$ \cite{Aubin:2008wk}, and $0.173(36)\GeV^4$ at $a=0.09\fm$ \cite{Aubin:2008wk}. It is also calculated for overlap valence and clover sea which yields $\Delta_{mix} = 0.35(14)/0.55(23)\GeV^4$ at $m_{\pi} = 190/300\MeV$ and $a=0.09\fm$~\cite{Durr:2007ef}.~This means that for a valence pion of $300\MeV$, $\Delta_{mix}$ produces, at $a=0.12\fm$, a shift of $\sim$ $110-240\MeV$ and,
at $a =0.09\fm$, a shift of $\sim$ $55-153\MeV$ for these cases, which are substantial portions of
the valence pion mass. 

We have used valence overlap fermions on the 2+1-flavor DWF sea to study hadron 
spectroscopy~\cite{Li:2010pw,Dong:2009wk,Mathur:2010ed,Gong:2011nr}.  In this work, we calculate 
$\Delta_{mix}$ for such a mixed action approach which is needed for chiral extrapolation
in MAPQ$\chi$PT.

\section{Calculation Details}
The $\Delta_{mix}$ parameter is calculated on four ensembles of the $2+1$-flavor domain wall fermion gauge configurations \cite{Allton:2008pn, Mawhinney:2009jy}. Two different lattice spacings were used to study the dependence on the cutoff. In addition, we also used multiple sea masses, for the $32^3 \times 64$ lattices, to study the sea quark mass dependence of $\Delta_{mix}$.   Details of the ensembles are listed in Table \ref{tab:ensembles}.
\begin{table}[htdp]
\begin{center}
\begin{tabular}{|c|c|c|c|}
\hline
Lattice Size& $a^{-1}(\GeV)$& $a m_l$& $ m_{ss}(\MeV)$\\ 
\hline
\hline
$24^3 \times 64$ & 1.73(3) & 0.005 & 329(1)\\
$32^3 \times 64$ & 2.32(3) & 0.004 & 298(1)\\
$32^3 \times 64$ & 2.32(3) & 0.006 & 350(2)\\
$32^3 \times 64$ & 2.32(3) & 0.008 & 399(1)\\
\hline
\end{tabular}
\end{center}
\caption{Details of the DWF ensembles used in this work.}
\label{tab:ensembles}
\end{table}

There are different parameterization schemes \cite{Orginos:2007tw, Aubin:2008wk} one can use to relate $\Delta_{mix}$ to the quark and pseudoscalar masses from Eq.~(\ref{eq:dd}). In this paper we choose to parameterize $\Delta_{mix}$ as 
 \begin{equation}
\label{eq:dmix}
m_{vs}^2 -\frac{1}{2}m_{ss}^2 = B_{ov} m_v + a^2\Delta_{mix}.
\end{equation}
This is similar to the parameterization used in \cite{Aubin:2008wk}. The quantity $\delta m^2 \equiv m_{vs}^2 -\frac{1}{2}m_{ss}^2$, has a linear behavior in $m_v$ and can be calculated directly by computing only the pseudoscalar masses $m_{vs}$, and $m_{ss}$.  We see that in the regime where Eq.~(\ref{eq:dmix}) is valid, $\Delta_{mix}$ is equivalent to $\delta m^2$ in the limit $ m_v \rightarrow 0$.

In this work we used an overlap operator with a HYP-smeared kernel. This was shown to have better numerical properties~\cite{Li:2010pw}. We calculated the masses $m_{vs}$ and $m_{ss}$ using 50 configurations for each ensemble.  Recall from Section I that $m_{ss}$ requires the propagators for the sea quark mass which were computed with DWF. $m_{vs}$ needs the propagators for both overlap and DWF. The DWF propagators were made available by LHPC.  The overlap propagators were computed using a polynomial approximation \cite{Alexandru:2011sc} to the matrix sign function. They  were used to compute $14-16$ values of $m_{vs}$. A multi-shifted version of the conjugate gradient algorithm \cite{Jegerlehner:1996pm}  was implemented for the overlap propagator calculation to compute all masses at once. To accelerate the inversions for the overlap propagators we employed a deflation technique which has been seen to speed up the calculation significantly~\cite{Li:2010pw}.  

We fitted correlators using single state exponential fits to extract the pion masses. The fitting windows were adjusted to get a reasonable $\chi^2/dof$ and are the same for all the masses in each ensemble. For the $24^3 \times 64$ ensemble the details are presented in Table \ref{tab:pfit1}.  For one of the sea quark mass in the  $32^3 \times 64$ ensemble, the results are presented in Table \ref{tab:pfit2}. 
\begin{table}[htdp]
\begin{center}
\caption{Pion mass fitting details for the $24^3 \times 64$ lattice. $m_{ss}  \simeq 329\MeV$.}
\begin{tabular}{|c|c|c|c|c|}
\hline
$a m_{v}$& $m_{vs}(\MeV)$  & $\chi^2_{vs}/dof$ & $m_{vv}(\MeV)$  &$\chi^2_{vv}/dof$ \\ 
\hline
\hline
0.0014& 274(3) & 1.0 &122(4) & 1.8 \\ 
0.0027& 278(2) & 1.1 &154(2) & 1.4 \\ 
0.0046& 287(2) & 1.2 &188(2) & 1.6 \\ 
0.0081& 305(2) & 1.2 &242(2) & 1.6 \\ 
0.0102& 315(2) & 1.2 &270(2) & 1.4 \\ 
0.0135& 331(2) & 1.1 &308(2) & 1.3 \\ 
0.0153& 339(1) & 1.1 &327(2) & 1.3 \\ 
0.0160& 342(1) & 1.1 &334(2) & 1.3 \\ 
0.0172& 347(1) & 1.1 &346(2) & 1.3 \\ 
0.0243& 378(1) & 1.2 &409(1) & 1.6 \\ 
0.0290& 397(1) & 1.2 &445(1) & 1.7 \\ 
0.0365& 426(1) & 1.2 &498(1) & 1.5 \\ 
0.0434& 451(1) & 1.2 &542(1) & 1.3 \\ 
0.0489& 471(1) & 1.2 &576(1) & 1.2 \\ 
0.0670& 531(1) & 1.0 &677(1) & 1.3 \\ 
0.0710& 543(1) & 1.0 &698(1) & 1.4 \\ 
\hline
\hline
\end{tabular}
\label{tab:pfit1}
\end{center}
\end{table}
\begin{table}[htdp]
\begin{center}
\caption{Pion mass fitting details for the $32^3 \times 64$ lattice for $a m_l = 0.004$.  $m_{ss}  \simeq 298\MeV$.}
\begin{tabular}{|c|c|c|c|c|}
\hline
$a m_{v}$& $m_{vs}(\MeV)$  & $ \chi^2_{vs}/dof$ & $m_{vv}(\MeV)$  &$\chi^2_{vv}/dof$ \\ 
\hline
\hline
0.0007& 237(3)&1.0& 126(10)&1.5\\
0.0015& 238(2)&1.0& 136(7)& 1.2\\
0.0025& 246(2)&1.0& 160(4)& 1.2\\
0.0035& 254(2)&1.0& 184(3)& 1.2\\
0.0046& 264(2)&1.0& 209(3)& 1.2\\
0.0059& 274(2)&1.0& 235(3)& 1.2\\
0.0068& 282(2)&1.0& 253(3)& 1.3\\
0.0076& 289(2)&1.0& 268(3)& 1.4\\
0.0089& 299(2)&1.0& 288(3)& 1.6\\
0.0112& 317(2)&1.1& 324(4)& 1.0\\
0.0129& 329(2)&1.1& 337(5)& 1.0\\
0.0152& 344(2)&1.2& 365(5)& 1.3\\
0.0180& 362(2)&1.3& 396(6)& 1.5\\
0.0240& 399(3)&1.6& 454(9)& 1.9\\
\hline
\hline
\end{tabular}
\label{tab:pfit2}
\end{center}
\end{table}%


\section{Fitting strategies and Results}

\begin{figure*}[t]
\includegraphics[width= 3.4in]{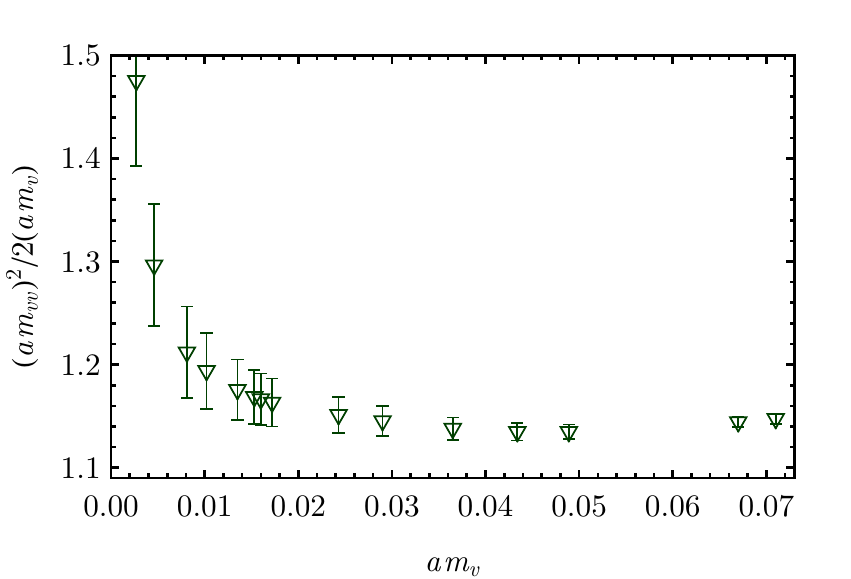}
\includegraphics[width= 3.4in]{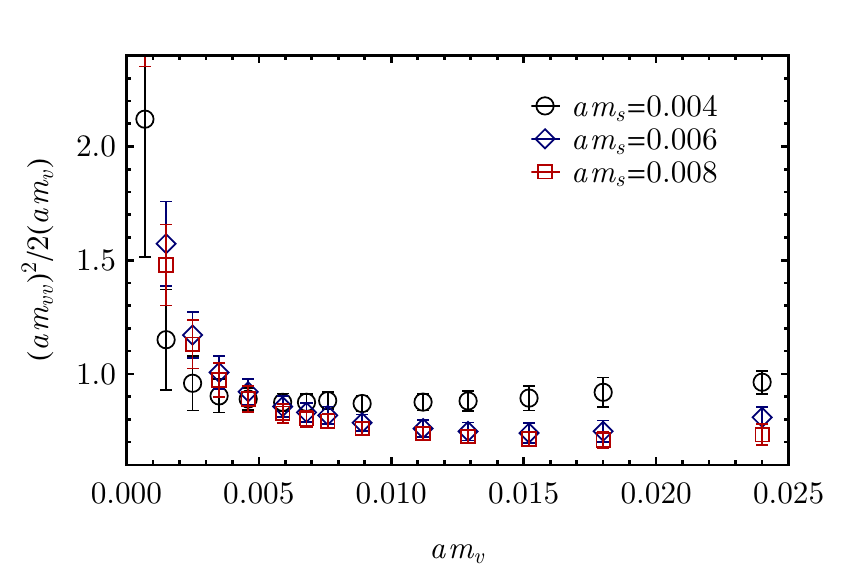}
\caption{$(a m_{vv})^2/2(a m_v)$ as a function of $a m_v$ for
the $24^3 \times 64$ ensemble (left) and 
the three $32^3 \times 64$ ensembles (right).}
\label{fitregion}
\end{figure*}

Extracting $\Delta_{mix}$ requires fits to squares of multiple pion propagators with different valence quark masses; the resulting 
masses will be correlated since they all come from a single ensemble. 
To compute the covariance matrix we would need to perform an augmented $\chi^2$-fit involving the pion propagators for all valence quark masses simultaneously.  This is not numerically stable due to the large number of parameters in the model.  We thus have to use an alternative procedure to take into account these correlations. To gauge the systematic errors introduced by our choice of fitting method we will use multiple fitting procedures. In this section we will describe three different fitting strategies. These methods differ in the way we account for correlations among the different valence masses.
 
In method I we follow the standard jackknife philosophy by defining a $\Delta_{mix}$ estimator directly in terms of the raw pion propagators; the error is then determined by the variance over the jackknife ensemble. In method II we use the jackknife procedure to estimate the $\delta m^2$ covariance matrix and then compute $\Delta_{mix}$ using a standard correlated fit~\cite{Luscher:2010ae}. These two procedures are used for both the $24^3 \times 64$ and $32^3 \times 64$ ensembles. However, since there are three sea quark masses for the $32^3 \times 64$ volume, we can perform an {\em uncorrelated} fit using $\delta m^2$ computed on the three independent ensembles---this is our method III. From the different methods we can obtain a systematic uncertainty for our results.   

In each method we performed a binned jackknife analysis of $\delta m^2 =m_{vs}^2 -\frac{1}{2}m_{ss}^2$. Because we performed four inversions with four different point sources per configuration we binned in units of four and eight. This allowed us to drop one or two whole configurations in each jackknife subsample. We find the uncertainties in both cases to be of comparable size, indicating that autocorrelation is negligible.  

The first step in our fitting is to determine a range of quark masses where the tree-level relation between $m^2_{vv}$ and $m_v$ holds so that we can use Eq.~(\ref{eq:dd}). Below this range, one expects to see chiral logs including partially quenched logs and other non-linear $m_v$ dependence from the NLO in $\chi$PT. Above this range, tree-level $\chi$PT is not expected to be valid.  We do this by plotting $(a m_{vv})^2 /2(a m_v)$ as a function of $a m_v$ as shown in Fig. \ref{fitregion}. We choose the fitting range in the region where the ratio $(a m_{vv})^2/2 (a m_v)$ is fairly flat; these ranges are tabulated in Table \ref{dmix:fitrange} and are used in all three fitting methods.  We note that in the range we are fitting, $m_{\pi}L > 4$ for both $m_{vs}$ and $m_{vv}$ so that the volume dependence is expected to be small.

\begin{table}[b]
\begin{center}
 \begin{tabular}{|c|c|c|}
  \hline
   Lattice&$a m_l$& $a m_v $ fit range\\
   \hline
   \hline
   $24^364$&0.005&0.0243 - 0.0489 \\
   $32^364$&0.004&0.0112 - 0.0240\\
   $32^364$&0.006&0.0112 - 0.0240 \\
   $32^364$&0.008&0.0112 - 0.0240\\
   \hline
   \hline
\end{tabular}
\end{center}
\caption{Range of quark masses, $a m_v$, used in the fitting procedures to extract $\Delta_{mix}$ via Eq.~(\ref{eq:dd}).}
\label{dmix:fitrange}
\end{table}

\begin{figure*}[t]
\includegraphics[width= 3.4in]{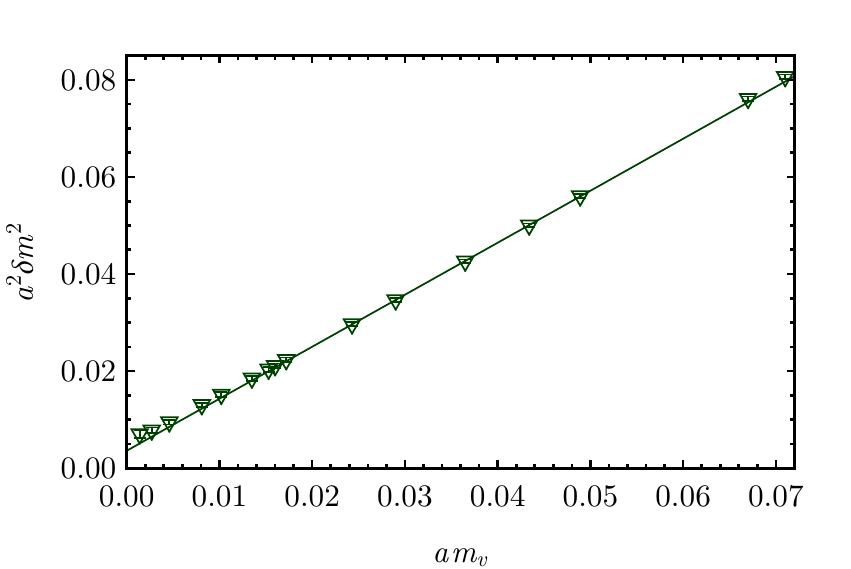}
\includegraphics[width= 3.4in]{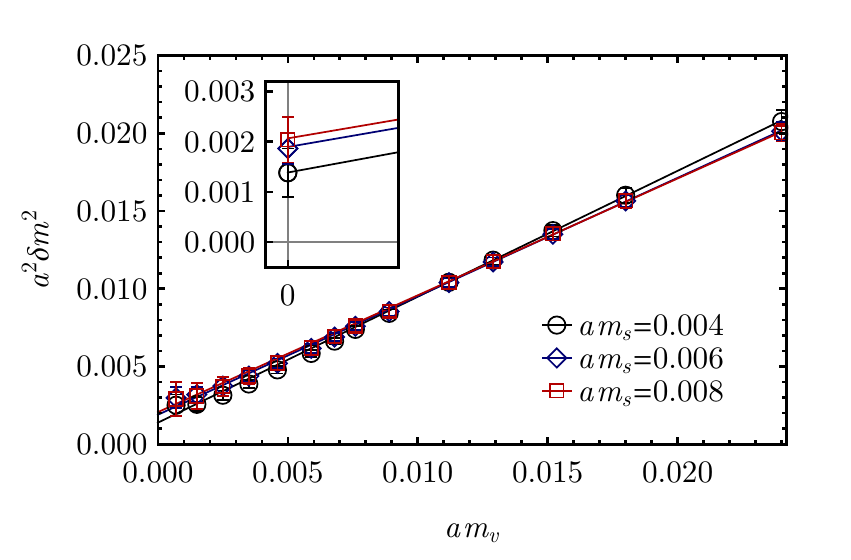}
\caption{Extracting $\Delta_{mix}$ from a linear extrapolation of $\delta m^2$ for the $24^3 \times 64$ ensemble (left) 
and the three $32^3 \times 64$ ensembles (right). For the $32^3 \times 64$ plot the inset shows the intercept of the 
fit and the error bars of the extracted values of $\Delta_{mix}$.}
\label{dmixfit}
\end{figure*}

\subsection{Method I:  weighted averaging}
In this method we used a weighted linear fit to extract the value of $\Delta_{mix}$ for each bin.  The weights, $\sigma_{\delta m^2}^2$, are given by 
\begin{equation}
\sigma_{\delta m^2}^2 = 4 m_{vs}^2\sigma_{ m_{vs}}^2 + m_{ss}^2 \sigma_{m_{ss}}^2,
\label{dmix:err1}
\end{equation}
where $\sigma_{m_{vs}}$, and  $\sigma_{m_{ss}}$ are the uncertainties of the mixed and DWF pion masses respectively. Eq.~(\ref{dmix:err1}) was derived using the standard error propagation formula neglecting correlation between $m_{vs}$ and $m_{ss}$. The cross-correlations will be accounted for by the external jackknife procedure. By using weighted fitting, less importance is given to data points with larger uncertainties.  After fitting each bin we then have a jackknife ensemble \{$\Delta_{mix}$\}. We use square brackets, [~], to indicate a particular jackknife sample and angle brackets, $\langle ~\rangle$, to denote jackknife averages.  For the case of $\Delta_{mix}$, its jackknife average and the corresponding uncertainty is given by
\begin{eqnarray}
\langle \Delta_{mix} \rangle &=& \frac{1}{N}\sum_{k=1}^{N}\left[ \Delta_{mix}\right]_k \nonumber \\
\sigma_{\Delta_{mix}} &=& \sqrt{(N-1)\left(\langle \Delta_{mix}^2\rangle-\langle \Delta_{mix}\rangle^2 \right)}\nonumber. \\
\end{eqnarray}

The extracted values of $\Delta_{mix}$, using this method, are presented in Table \ref{dmix:method1}. We also list the corresponding values of $B_{ov}$.  Figure \ref{dmixfit} shows $a^2 \delta m^2$ as a function of $a m_{v}$ and the corresponding linear fit.

\begin{table}[b]
\begin{center}
 \begin{tabular}{|c|c|c|c|c|c|}
 \hline
   \multirow{2}{*}{Lattice}&\multirow{2}{*}{$a m_l$}& \multicolumn{2}{|c|}{$\Delta_{mix}(\GeV^4)$}& \multicolumn{2}{|c|}{$B_{ov}$(\GeV)}\\
   &&\multicolumn{1}{c}{I}&II&\multicolumn{1}{c}{I}&II\\
   \hline
   \hline
   $24^364$&0.005&0.032(6)  & 0.028(5)  &1.85(2)  &1.88(2) \\
   $32^364$&0.004&0.040(14)& 0.025(9)  &1.88(10)&1.95(6)\\
   $32^364$&0.006&0.054(9)  & 0.050(8)  &1.81(5)  &1.81(5)\\
   $32^364$&0.008&0.059(13)& 0.063(13)&1.74(6)  &1.73(6) \\
   \hline
   \hline
\end{tabular}
\end{center}
\caption{Extracted values of $\Delta_{mix}$ and $B_{ov}$ using fitting methods I and II.}
\label{dmix:method1}
\end{table}


\subsection{Method II: covariance matrix}
In our second method we perform a correlated valence-quark mass fit to the function $\delta m^2  = B_{ov} m_v + a^2 \Delta_{mix}$. The central values are taken to be the jackknife averages,
\begin{equation}
\langle(\delta m^2)_i \rangle = \frac{1}{N}\sum_{k=1}^N\left [(\delta m^2)_i\right]_k \,.
\end{equation}
We minimize the correlated $\chi^2$-function,
\begin{eqnarray*}
\chi^2& =& \sum_{i,j} \left(\langle(\delta m^2)_i \rangle - (\delta m^2)_i\right)C_{i,j}^{-1} \nonumber\\
&\times&\left(\ \langle(\delta m^2)_j \rangle -(\delta m^2)_j\right) \,, \nonumber \\
\end{eqnarray*}
with the covariance matrix, $C_{i,j}$, given by the jackknife estimate
\begin{eqnarray}
C_{ij} &=& \frac{(N-1) } {N} \sum_{k=1}^N \left(\left[(\delta m^2)_i\right]_k - \langle(\delta m^2)_i \rangle \right) \nonumber \\
&\times& \left(\left[(\delta m^2)_j\right]_k - \langle (\delta m^2)_j \rangle\right) \,. \nonumber \\
\label{covmat}
\end{eqnarray}
The subscripts $i$ and $j$ index the valence-quark mass. The factor $(N-1)/N$ differs from the usual definition of the covariance matrix to account for the correlation of the jackknife samples~\cite{Luscher:2010ae}. The fit uncertainties are obtained by constructing the standard Hessian matrix. Results of this method are tabulated in Table \ref{dmix:method1}.
\subsection{Method III: uncorrelated fitting}
The methods described in the two previous sections closely parallel the method performed by \cite{Aubin:2008wk} in which for each lattice ensemble a string of partially quenched meson masses are calculated; from them we extract $\Delta_{mix}$.  In this section we perform a fit on the three $32^3 \times 64$ ensembles based on the value of $\delta m^2$ measured at the point where the valence and sea quark masses match \footnote{As mentioned in \cite{Aubin:2008wk} one can never get rid of partial quenching effects even in the case where the pseudoscalar mesons of both the DWF and overlap actions match. This is because discretization errors are different for the two actions.}.  There are no cross-correlations among the masses since the ensembles are independent. 

The value of $\delta m^2$ is computed for the pion mass that most closely satisfies the condition $m_{ss} \approx m_{vv}$. Because it is not  easy to match a priori the pseudoscalar masses and it is expensive to regenerate overlap propagators with different masses, we performed interpolation among the existing data points to obtain a better approximation of where $m_{ss}$ and $m_{vv}$ match.

\begin{figure}[t]
\includegraphics[width= 3.4in]{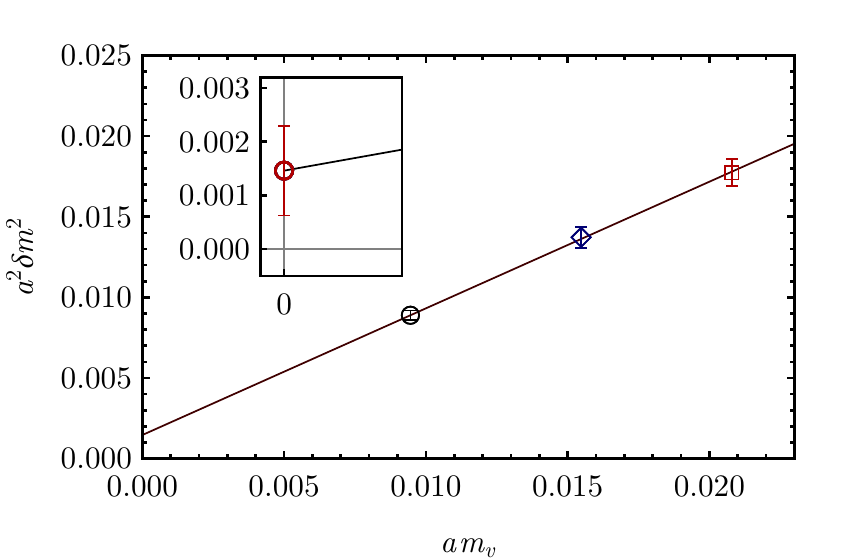} 
\caption{Determining $\Delta_{mix}$ by performing the linear extrapolation as in  Fig.~\ref{dmixfit}.  Each point corresponds to one of the $32^3 \times 64$ lattices where $m_{ss} \approx m_{vv}$. A magnified view of the extrapolation in the neighborhood of the chiral limit, along with the extrapolated value, is shown in the inset.}
\label{dmix:32method3}
\end{figure}

We do an uncorrelated $\chi^2$-fit with the error bars obtained by a jackknife procedure. Figure \ref{dmix:32method3} shows the data points and the fit results. For this case we find $\Delta_{mix} = 0.042(24)\GeV^4$, and $B_{ov} = 1.82(16)\GeV$. 

\section{Discussion}


As a first step, we look at the lattice spacing dependence. We compare the two ensembles with $m_{ss}$ close to $300\MeV$.  
These correspond to $a m_l = 0.005$ and $a m_l = 0.004$ with $a = 0.114\fm$ and $a=0.085\fm$ respectively.
We average the central values and the errors from the two fitting methods, separately for each lattice spacing. 
We find at $a = 0.114\fm$ $\Delta_{mix} = 0.030(6)\GeV^4$ and for $a = 0.085\fm$ $\Delta_{mix} = 0.033(12)\GeV^4$. 
We see that the lattice spacing dependence is smaller than our errors; this indicates that $\Delta_{mix}$ is capturing 
the dominant lattice artifact for the parameters used in this study.  

\begin{figure}[t]
\includegraphics[width= 3.4in]{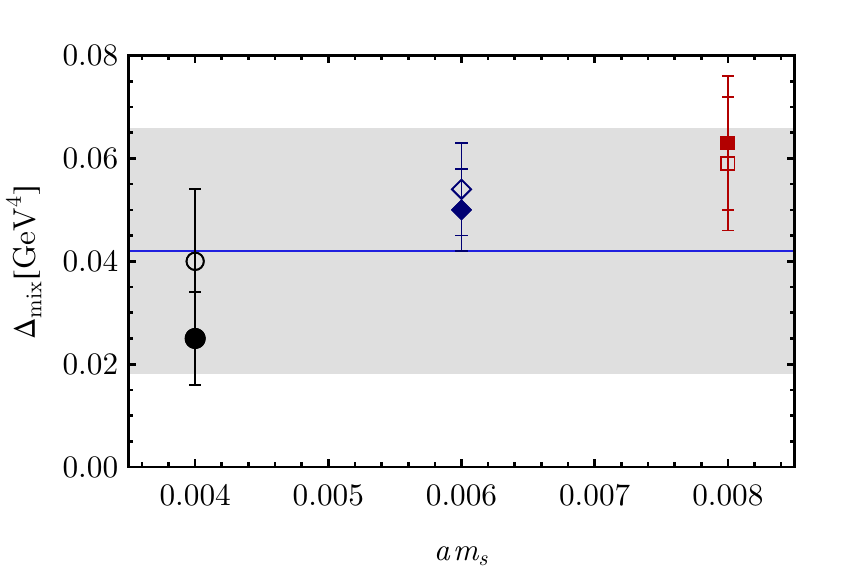} 
\caption{$\Delta_{mix}$ for $ a= 0.085\fm$. The empty symbols indicate the results of method I and the full symbols are from method II. The continuous line and the shaded region are the results of the chiral extrapolation in method III.}
\label{dmix:32method4}
\end{figure}

To discuss the sea quark dependence of $\Delta_{mix}$ we plot in Fig.~\ref{dmix:32method4} the results of 
methods~I, II, and III for the ensembles with $a = 0.085\fm$.  We note that the results are consistent within 
two sigma. This is consistent with LO MAPQ$\chi$PT in Eq.~(\ref{eq:dd}) where $\Delta_{mix}$ is a low energy 
constant, independent of the valence and sea masses.

We now combine the $\Delta_{mix}$ values extracted from the two lightest sea mass ensembles to produce our final result.
We use these ensembles because the LO MAPQ$\chi$PT is expected to describe the data better at lower sea quark masses.
For each of the fitting methods we combine the results of the $a = 0.114\fm$ and $a=0.085\fm$ ensembles.
Since the two ensembles are statistically independent, it is straightforward to combine these results: for method~I we get 
$\Delta_{mix} = 0.033(6)\GeV^4$ and for method~II we get $\Delta_{mix} = 0.027(5)\GeV^4$. We can now use the values
determined using these two methods to estimate the systematic fitting errors. 
We quote the final result with two uncertainties, the first statistical and the second associated with fitting systematics.
The central value and the statistical error are taken to be the average of the results from methods~I and II. 
The systematic error is the standard deviation of the results from these two methods. 
We get $\Delta_{mix}=0.030(6)(5)\GeV^4$.


Table \ref{dmix:others} lists calculated values of $\Delta_{mix}$ for
pion masses close to $300\MeV$ using different mixed actions.  We
notice that our values of $\Delta_{mix}$ are  significantly smaller
than overlap on clover or DWF on Asqtad for comparable lattice spacings and pion masses. 
For both $m_{vv}$ and $m_{ss}$ at $300\MeV$, our calculated $\Delta_{mix}$ will
shift the pion mass up by $10$ and $16\MeV$ for the $32^3 \times 64$ lattice at
$a = 0.085\fm$ and $24^3 \times 64$ lattice at $a = 0.114\fm$, respectively.
These are substantially smaller than the corresponding $55-153\MeV$ and
$110-240\MeV$ shifts that we mentioned in Sec. I.
The value of $\Delta_{mix}$ is significantly smaller in our case probably because 
the sea and valence fermion actions are similar.  Both of them are approximations of 
the matrix sign function, but with different kernels.

\begin{table}[htbp]
\begin{center}
\begin{tabular}{|l|c|c|c|}
\hline
~~~~Mixed Action& Ref. & $a\fm$&$\Delta_{mix}(\GeV^4)$\\
\hline
\hline
DWF on staggered& \cite{Orginos:2007tw}&0.125& 0.249(6)\\
DWF on staggered&\cite{Aubin:2008wk}& 0.12& 0.211(16)\\
DWF on staggered& \cite{Aubin:2008wk}& 0.09& 0.173(36)\\
overlap on clover& \cite{Durr:2007ef}&0.09&0.55(23)\\
overlap on DWF    & this work & 0.114 & 0.030(6)\\
overlap on DWF    & this work & 0.085& 0.033(12)\\
\hline
\end{tabular}
\end{center}
\caption{$\Delta_{mix}$ values for DWF valence quarks on staggered sea quarks for pion mass at $300\MeV$.}
\label{dmix:others}
\end{table}

The previous calculation \cite{Li:2010pw} of $\Delta_{mix}$, for
overlap on the DWF sea, has roughly the same value but the sign is
negative. That calculation measured $\Delta_{mix}$ by examining the
states that wrap around the time boundary.  This indirect 
method is less reliable and the errors are large. 

\section{Conclusion}
We calculated the additive mixed action pseudoscalar meson mass parameter,
$\Delta_{mix}$, for the case of valence overlap fermions on a DWF sea.
$\Delta_{mix}$ is significant because it enters
into mixed action partially quenched chiral perturbation theory (MAPQ$\chi$PT) for
chiral extrapolation of many low energy observables.

Two different lattice spacings were used to examine the cut-off behavior.
For a pion mass close to $300\MeV$ we find $\Delta_{mix}=0.030(6)\GeV^4$ at $a =0.114\fm$ and 
$\Delta_{mix}=0.033(12)\GeV^4$ at $a =0.085\fm$.  They are the same within errors. 
Our calculated $\Delta_{mix}$ will
shift the pion mass up by $10$ and $16\MeV$ for the $32^3 \times 64$ lattice at
$a = 0.085\fm$ and $24^3 \times 64$ lattice at $a = 0.114\fm$, respectively.
We studied the sea quark mass dependence of $\Delta_{mix}$ at $a = 0.085\fm$ and
we find that they agree within two sigma. Combining the results extracted from
the ensembles with the lightest sea quarks, we get $\Delta_{mix}=0.030(6)(5)\GeV^4$,
where the first error is statistical and the second is the systematic error
associated with the fitting method.
 
When compared to previous mixed action studies, DWF on staggered or overlap on clover, the values of $\Delta_{mix}$ of overlap on  DWF
are almost an order of magnitude smaller. This is most likely due to the fact that the sea and valence fermions
are similar. 

\begin{acknowledgements}
We would like to thank J. Negele, A. Pochinsky, M. Engelhardt, and
LHPC for generously making available the DWF propagators for all the
ensembles used in this paper. This work is supported in part by
U.S. Department of Energy grants DE-FG05-84ER40154,
DE-FG02-95ER40907, GW IMPACT collaboration, DE-FG02-05ER41368, and DST-SR/S2/RJN-19/2007, India.
\end{acknowledgements}

\newpage
\bibliographystyle{jhep-3}
\bibliography{my-references}
\end{document}